\documentclass[preprint,aps,preprintnumbers,amsmath,amssymb,superscriptaddress]{revtex4}

\usepackage{graphicx}
\usepackage{dcolumn}
\usepackage{bm}

\begin{document}


\title{Macroscopic quantum computation using Bose-Einstein condensates}
\author{Tim Byrnes}
\affiliation{National Institute of Informatics, 2-1-2
Hitotsubashi, Chiyoda-ku, Tokyo 101-8430, Japan}

\author{Kai Wen}
\affiliation{E. L. Ginzton Laboratory, Stanford University, Stanford, CA 94305}

\author{Yoshihisa Yamamoto}
\affiliation{National Institute of Informatics, 2-1-2
Hitotsubashi, Chiyoda-ku, Tokyo 101-8430, Japan}
\affiliation{E.
L. Ginzton Laboratory, Stanford University, Stanford, CA 94305}
\date{\today}

\begin{abstract}
Quantum computation using qubits made of two component Bose-Einstein condensates (BECs) is analysed. The use 
of BECs allows for an increase of energy scales via bosonic enhancement, resulting in gate operations that can be performed at a macroscopically large energy scale. The large energy scale of the gate operations results in quantum algorithms that may be executed at a time reduced by a factor of $N$, where $ N $ is the number of bosons per qubit. The encoding of the qubits allows for no intrinsic penalty on decoherence times. We illustrate the scheme by an application to Deutsch's and Grover's algorithms. 
\end{abstract}

\maketitle

Recent advances in semiconductor technology has allowed the realization of single qubit decoherence
rates as small as $ \sim 3 \times 10^{-6} $ \cite{press08,berezovsky08,press10} in spin quantum dots. 
Such decoherence rates are a result of the ability to optically control the qubit states using bright 
coherent pulses, resulting in ultrafast picosecond gates.  This allows for a vastly increased number of 
gates allowable within a limited decoherence time $ T_2 $.  One of the key ingredients for the success of these ultrafast gates originates from the effect of bosonic enhancement. The Rabi frequency between two levels is given by $ \Omega = \mu E /\hbar $, where $ \mu $ is the optical transition dipole matrix element and $ E $ is the electric field amplitude.  Since $ E $ is typically a macroscopically large quantity, in principle it is possible increase the energy scale (and thereby decrease the time scale) of the single qubit gate by simply increasing the laser power. This bosonic enhancement is ubiquitous in 
many areas of quantum optics and condensed matter physics, in phenomena such as laser physics, resonance flourescence and Bose-Einstein condensation (BEC), where quantum coherence is ``amplified'' when a large number of bosons are present in the system. Our aim in this work is to bring quantum computation to a macroscopically large energy scale, much in the same way that lasers bring the quantum coherence of light to a macroscopic level. 

Although the energy scale for single qubit rotations may be amplified in the way described above, for two qubit gates, a straightforward extension of this idea is typically problematic. For example, consider 
a two qubit gate connected via a photonic quantum bus \cite{pellizzari95,zheng00}. Introducing a large population of photons into the quantum bus typically results in additional sources of decoherence, due to the leakage of the photons out of the system. Thus typically very weak excitations are assumed in the quantum bus to minimize such effects. 
Here we describe an alternate way of increasing the energy scale of the interactions between qubits by populating each qubit with a large number of identical bosons. The 
large number of bosons on each qubit possesses a bosonic enhancement that increases the energy scale of the control Hamiltonian, which equates to faster control operations.

\begin{figure}
\scalebox{0.6}{\includegraphics{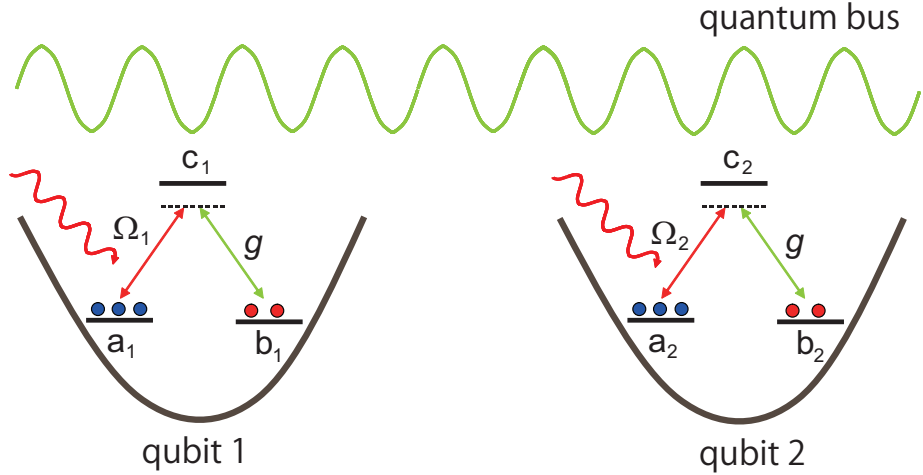}}
\caption{\label{fig1} Two bosonic qubits mediated by a quantum bus. The quantum bus couples transitions between levels $ b_n $ and $ c_n $ with energy $ g $. Individual pulses coupling levels $ a_n $ and $c_n$ with frequency $ \Omega_n $ create an adiabatic passage between levels $ a_n $ and $ b_n $. }
\end{figure}

Consider a bosonic system with two independent degrees of freedom \cite{sorensen00}, such as two hyperfine levels of an atomic Bose-Einstein condensate (BEC) or spin polarization states of exciton-polaritons \cite{deng10}. Such two component BECs have been realized and controlled in atom chip systems, with the demonstration of spin squeezing \cite{riedel10}. Denote the bosonic annihilation operators of the two states as $ a $ and $ b $, obeying commutation relations $ [a,a^\dagger]= [b,b^\dagger]= 1 $ \cite{li09}.
We then encode each qubit in a state
\begin{align}
\label{singlequbitstate}
|\alpha, \beta \rangle \rangle \equiv \frac{1}{\sqrt{N!}} \left( \alpha a^\dagger + \beta b^\dagger \right)^N |0 \rangle ,
\end{align}
where $ \alpha $ and $ \beta $ are arbitrary complex numbers satisfying $ |\alpha |^2 + | \beta |^2 = 1 $ 
(double brackets are used to denote the bosonic qubit states).  We consider the boson number $ N = a^\dagger a + b^\dagger b $ to be a conserved number. 

The state $ |\alpha, \beta \rangle \rangle $ can be manipulated using Schwinger boson (Stokes operators) operators $ S^x = a^\dagger b + b^\dagger a $, $ S^y = -i a^\dagger b + i b^\dagger a $, $ S^z = a^\dagger a - b^\dagger b $, which satisfy the usual spin commutation relations $ [S_i,S_j] = 2i \epsilon_{ijk} S_k $, where 
$ \epsilon_{ijk} $ is the Levi-Civita antisymmetric tensor.  In the spin language, (\ref{singlequbitstate}) forms a $ S= N/2 $ representation of the SU(2) group.  Single qubit rotations can be performed in a completely analogous fashion to regular qubits.  For example, rotations around the z-axis of the Bloch sphere can be peformed by an evolution
\begin{align}
e^{-i S^z t} |\alpha, \beta \rangle \rangle & = 
\frac{1}{\sqrt{N!}} \sum_{k=0}^N {N \choose k} ( \alpha a^\dagger e^{-it})^k ( \beta b^\dagger e^{it})^{N-k} |0 \rangle \nonumber \\
& = |\alpha e^{-it}, \beta e^{it}\rangle \rangle 
\end{align}
Similar rotations may be performed around any axis by an application of
\begin{align}
H_{\mbox{\tiny qub}} = \bm{n} \cdot \bm{S} = n_x S^x + n_y S^y +n_z S^z 
\end{align}
where $ \bm{n} = (n_x,n_y,n_z) $ is a unit vector. Expectation values of the spin are exactly analogous to that of a single spin, taking values
\begin{align}
\langle S^x \rangle & = N(\alpha^* \beta + \alpha \beta^*) \nonumber \\
\langle S^y \rangle & = N(-i \alpha^* \beta + i \alpha \beta^*) \nonumber \\
\langle S^z \rangle & = N( | \alpha |^2 - | \beta |^2 ) .
\end{align}
The variance of the spins however diminish relative to the maximum amplitudes
\begin{align}
\label{variance}
(\langle (S^z)^2 \rangle - \langle S^z \rangle^2)/N^2 = 4 |\alpha \beta |^2 /N
\end{align}
in accordance to widespread notion that for $ N \rightarrow \infty $ the spins approach classical variables. 

Two qubit interactions can be performed using a generalization of the methods in Ref. \cite{zheng00}. 
Consider an interaction Hamiltonian 
\begin{align}
\label{qubushamiltonian}
H_{\mbox{\tiny bus}} = \frac{\omega_0}{2} \sum_{n=1,2} F^z_n + \omega p^\dagger p + g  \sum_{n=1,2} \left[ F^-_n p^\dagger + F^+_n p \right] ,
\end{align}
where $ F^z =c^\dagger c - b^\dagger b$, $ F^+ = c^\dagger b  $, $ \omega_0 $ is the transition energy, and $ p $ is the photon annihilation operator (we set $ \hbar=1 $). Assuming a large detuning $ \Delta = \omega_0 - \omega \gg g \sqrt{N} $, we may adiabatically eliminate the photons from the bus by assuming $ p^\dagger p =0 $ and we obtain an 
effective Hamiltonian $ H_{\mbox{\tiny bus}} \approx \frac{g^2}{\Delta} \left( F^+_1 F^-_2 + F^-_1 F^+_2 \right) $. Now consider a further detuned single qubit transition according to $ H_{\mbox{\tiny pul}} = \Omega(t) \sum_{n=1,2} \left[  c_n^\dagger a_n + \mbox{H.c} \right]$.  After adiabatic elimination of level $ c $ by assuming $ c^\dagger_n c_n = 0 $, we obtain
\begin{align}
\label{expandedinteraction}
H_{\mbox{\tiny int}} \approx \frac{g^2 \Omega (t)}{\Delta^2} \left( S^+_1 S^-_2 + S^-_1 S^+_2 \right) .
\end{align}
The energy scale of the interaction term is then $ \sim O( g^2 \Omega N^2/\Delta^2) $, since the spin operators themselves are of order $ S^z \sim O(N) $.  Therefore, by replacing the qubits by the collective states (\ref{singlequbitstate}) we have managed to create an interaction term with a boosted energy scale of $~ N^2 $. We note that alternative ways of producing two-qubit interactions are also possible, such as by using cold controlled collisions \cite{jaksch99}. 

The consequence of the boosted energy scale of the interaction can be observed by examining explicitly the state evolution of two qubits. Before showing this, let us note here that the combination of $ H_{\mbox{\tiny qub}} $ and $ H_{\mbox{\tiny int}}$ may be combined to form an arbitrary Hamiltonian involving spin operators according to universality arguments \cite{lloyd95}. By successive commutations an arbitrary product of spin Hamiltonians $ H = \prod_{n=1}^M (S^\alpha_n)^{m(n)} $ may be produced, where $ M $ is the total number of qubits, $ \alpha = x,y,z $, and $ m(n) = 0,1 $. Using this fact, we consider henceforth the interaction Hamiltonian $ H_{\mbox{\tiny int}} = S^z_1 S^z_2 $ instead of (\ref{expandedinteraction}), for the simplicity of the analysis.
As a simple illustration, let us perform the analogue of the maximally entangling operation 
\begin{align}
& e^{-i \sigma^z_1 \sigma^z_2 \frac{\pi}{4} } ( | \uparrow \rangle + | \downarrow \rangle ) ( | \uparrow \rangle + | \downarrow \rangle ) =  | + y \rangle | \uparrow \rangle  + | - y \rangle | \downarrow \rangle, \label{qubitentangler}
\end{align}
where $ | \pm y \rangle = e^{\mp i\frac{\pi}{4}} | \uparrow \rangle + e^{\pm i\frac{\pi}{4}} | \downarrow \rangle $. Starting from two unentangled qubits, we may apply $ H_{\mbox{\tiny int}} $ to obtain
\begin{align}
 e^{-i S^z_1 S^z_2 t} & | \frac{1}{\sqrt{2}}, \frac{1}{\sqrt{2}} \rangle \rangle | \frac{1}{\sqrt{2}}, \frac{1}{\sqrt{2}} \rangle \rangle = \frac{1}{N! 2^N} \sum_{k_1,k_2=0}^N {N \choose k_1} {N \choose k_2}   \nonumber \\
& (a_1^\dagger)^{k_1} (b_1^\dagger)^{N-k_1} ( a_2^\dagger)^{k_2} ( b_2^\dagger)^{N-k_2} e^{-i(N-2 k_1)(N-2 k_2)t} |0 \rangle  \nonumber \\
= & \frac{1}{\sqrt{2^N}} \sum_{k_2} \sqrt{N \choose k_2}  | \frac{e^{i(N-2 k_2)t}}{\sqrt{2}}  , \frac{e^{-i(N-2 k_2)t}}{\sqrt{2}}  \rangle \rangle |k_2 \rangle ,
\label{bosonqubitentanglement}
\end{align}
where in the second equality we have introduced eigenstates of the $ S^z $ operator 
$ |k \rangle = \frac{(a^\dagger)^k (b^\dagger)^{N-k}}{\sqrt{k!(N-k)!}} |0 \rangle $. For gate times equal to $ t = \pi/4N $ we obtain the analogous state to (\ref{qubitentangler}). For example, projecting qubit 2 onto the maximum $z$ eigenstates $ |k_2=0,N \rangle $ gives the states  $ | \frac{e^{\pm i\pi/4}}{\sqrt{2}}  , \frac{e^{\mp i \pi/4}}{\sqrt{2}} \rangle \rangle  $, which is the analogue of a Bell state for the bosonic qubits.

\begin{figure}
\scalebox{0.6}{\includegraphics{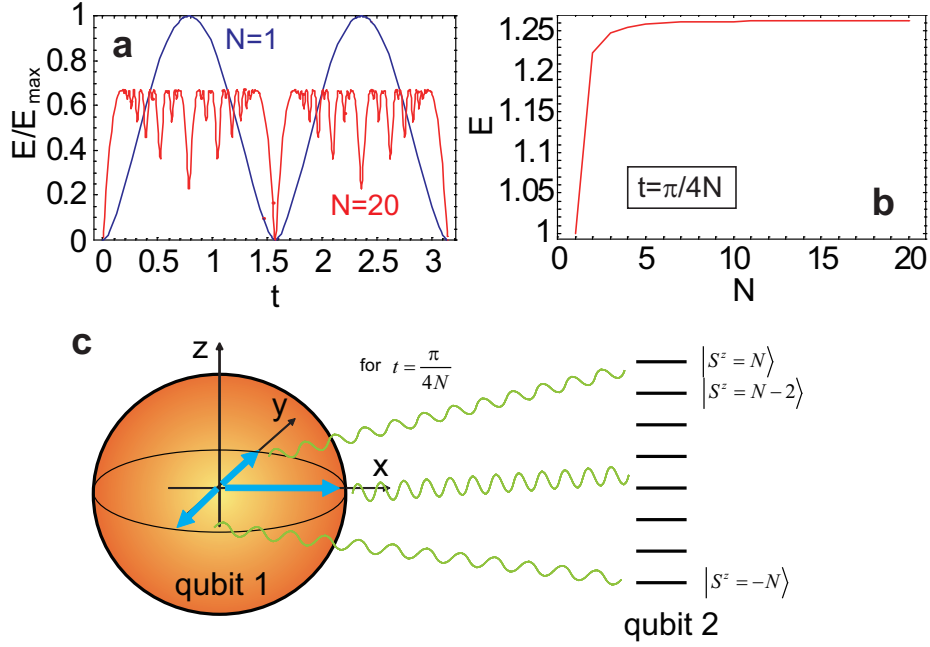}}
\caption{\label{fig2}
{\bf a} The entanglement normalized to the maximum entanglment ($E_{\mbox{\tiny max}} = \log_2 (N+1) $) between two bosonic qubits for the particle numbers as shown. {\bf b} Entanglement at a time $ t= \pi/4N $ for various boson numbers $ N $.  {\bf c} A schematic representation of the entangled state (\ref{bosonqubitentanglement}). }
\end{figure}

A visualization of the state (\ref{bosonqubitentanglement}) is shown in Figure \ref{fig2}c. For each $z$-eigenstate on qubit 2, there is a state $ | \frac{e^{i(N-2 k_2) \pi/4N}}{\sqrt{2}}  , \frac{e^{-i(N-2 k_2) \pi/4N}}{\sqrt{2}}  \rangle \rangle  $ on qubit 1 represented on the Bloch sphere entangled with it.  The states on qubit 1 are not orthonormal to each other except for the extremal states $ | \frac{e^{\pm i \pi/4}}{\sqrt{2}}  , \frac{e^{\mp i \pi/4}}{\sqrt{2}}  \rangle \rangle$, but the $z$-eigenstate on qubit 2 are of course orthonormal. Thus the type of entangled state is a continuous verson of the original qubit sequence (\ref{qubitentangler}), and has similarities to continuous variable formulations of quantum computing \cite{braunstein05}, although the class of states that are used here are quite different. We note here that the analogue of the CNOT gate can be produced by 
further evolving (\ref{bosonqubitentanglement}) with the Hamiltonian $ H = NS^z_1-NS^z_2+ N^2 $ for a time $ t = \pi/4N $ which gives
\begin{align}
\frac{1}{\sqrt{2^N}} \sum_{k_2} \sqrt{N \choose k_2}  | \frac{e^{-i \pi k_2/N}}{\sqrt{2}} , \frac{1}{\sqrt{2}} \rangle \rangle |k_2 \rangle .
\end{align}
Projecting onto the $ |k_2=0,N \rangle $ states, qubit 1 is left in the states $ | \pm \frac{1}{\sqrt{2}} ,\frac{1}{\sqrt{2}} \rangle \rangle $ which is a CNOT operation with the target qubit in the $x$-basis for qubit 1.

The crucial point to notice in (\ref{bosonqubitentanglement}) is that a gate time of $ t=\pi/4N $ was required to produce this entangled state, in comparison to the standard qubit case of $ t=\pi/4 $.  The 
reduced gate time arises due to the boosted energy scale of the interaction Hamiltonian, which in turn
is due to bosonic amplification of the interaction. This has similarities with other proposals for decreasing cooling times by bosonic final state stimulation \cite{byrnes09}. 

Despite the widespread belief that for $ N \rightarrow \infty $ the spins approach classical variables according to (\ref{variance}), the entangling operation (\ref{bosonqubitentanglement}) generates genuine entanglement between the bosonic qubits.  As a measure of the entanglement, we plot the von Neumann entropy 
$ E = - \mbox{Tr} ( \rho \log_2 \rho ) $ \cite{nielsen00} in Figure \ref{fig2}a.  For the standard qubit case ($N=1$), the entropy reaches its maximal value at $ t = \pi/4 $ in accordance with (\ref{qubitentangler}).  For the bosonic qubit case there is an initial sharp rise, corresponding to the improvement in speed of the entangling operation, but later saturates to a non-maximal value due to the presence of the binomial factors in (\ref{bosonqubitentanglement}) biasing the states towards zero spin values.  In Figure \ref{fig2}b we show the amount of entanglement present at times $ t=\pi/4N $, corresponding to the analogous state to (\ref{qubitentangler}).  We see that at such times there is approximately the same amount of entanglement as for the $ N= 1 $ case as for large $ N $, reinforcing the 
intuition that the $ e^{-iS^z_1 S^z_2 \pi/4N} $ gate gives the bosonic analogy to the operation (\ref{qubitentangler}).

The purpose of boosting the energy scale of the Hamiltonian is defeated if decoherence times also decrease when the number of bosons are increased. Naively one might guess that a large number of bosons encoding a single qubit would be more susceptible to decoherence. This is however is not necessarily always true.  
As a simple model of dephasing and particle loss, consider the evolution of Lindblad form through the master equation
\begin{align}
\frac{d \rho}{dt} = - \Gamma \sum_n [A_n,[A_n,\rho]],
\end{align}
where $ \Gamma $ is the dephasing rate, and $ A_n $ is the coupling operator to the environment, taken to be $ S^z_n $ for dephasing and $ a_n, b_n $ for particle loss. For a multiple qubit system, the 
information in a general quantum state can be reconstructed by $ 4^M - 1 $ expectation values of $ (I_1,S^x_1,S^y_1,S^z_1) \otimes (I_2,S^x_2,S^y_2,S^z_2)  \dots \otimes (I_M,S^x_M,S^y_M,S^z_M) $ \cite{altepeter04}.  For the bosonic system, there are in general higher order correlations involving 
powers of operators beyond order one (e.g. $ (S^x_n)^2 $).  However, since we perform the bosonic mapping
in order to perform a simulation of the qubit system, the presence of such expectation values are
of no concern in our case.  

Examining the dephasing of the general correlation $ \langle \prod_n S^{\alpha(n)}_n \rangle $ where $ \alpha(n) = I,x,y,z $, we obtain the evolution equation $ d\langle \prod_n S^{\alpha(n)}_n \rangle/dt = -4 \Gamma K \langle \prod_n S^{\alpha(n)}_n \rangle $, which has a solution
\begin{align}
\langle  \prod_n S^{\alpha(n)}_n \rangle \propto \exp[-4 \Gamma K t] .
\end{align}
Here $ K $ is a constant which is at most equal to $ M $, depending the particular $ \alpha(n) $ chosen.  
This equation does not have any $ N $ dependence, and in fact behaves identically to the qubit case ($N=1$).  Similar results hold for particle loss. Physically this difference is due to the statistical independence of the dephasing and particle loss processes among the bosons.  
Given a qubit algorithm intended for two-level qubits, how does this translate in the bosonic system? We have found that for most applications, the procedure amounts to: (i) finding the sequence of Hamiltonians required for the algorithm, (ii) making the replacement $ \sigma^\alpha_n \rightarrow N S^\alpha_n $, $ \sigma^\alpha_n \sigma^\beta_m  \rightarrow S^\alpha_n S^\beta_m $, (iii) Evolving the same sequence of Hamiltonians for a reduced time $ t \rightarrow t/N $.  This approach is reasonable from the point of view that we are performing the same algorithm except that a higher representation of SU(2) is being used. Let us illustrate this procedure with two well-known quantum algorithms with speedups over classical algorithms.  

{\it Deutsch's algorithm} \cite{nielsen00}. We reformulate the standard qubit version ($N=1$) of the algorithm in the following form convenient for our purposes.  The oracle performing the function $ | x \rangle | y \rangle \rightarrow | x \rangle | f(x) \oplus y \rangle $ is assumed to be one of the four Hamiltonians $ H_D = \{ 0, 2\sigma^z_2,\sigma^z_1 \sigma^z_2 + \sigma^z_2 -1,-\sigma^z_1 \sigma^z_2 + \sigma^z_2 -1 \} $ and evolved for a time $ t=\pi/4 $, which correspond to the functions $ f(x) = \{ (0,0),(1,1),(0,1),(1,0) \} $ respectively. The initial state is assumed to be the state $ (\uparrow + \downarrow) \uparrow $, and a measurement of qubit 1 in the $ x$-basis distinguishes between constant and balanced functions via the results $ (\uparrow + \downarrow)$ and $ (\uparrow - \downarrow)$ respectively. 

This can be translated into the corresponding algorithm for bosonic qubits according to the following procedure.  The oracle is assumed to be one of the following Hamiltonians $ H_D = \{ 0, 2 N S^z_2,S^z_1 S^z_2 + N S^z_2 -N^2,-S^z_1 S^z_2 + N S^z_2 -N^2 \} $, and we prepare the initial state as $ | \frac{1}{\sqrt{2}} , \frac{1}{\sqrt{2}} \rangle \rangle | 1 , 0 \rangle \rangle $. After evolving the Hamiltonians for a time $ t = \pi/4N $, we obtain (up to an overall phase)
\begin{align}
e^{-i H_D \pi/4N } | \frac{1}{\sqrt{2}} , \frac{1}{\sqrt{2}} \rangle \rangle | 1 , 0 \rangle \rangle = 
| \frac{1}{\sqrt{2}} , \pm \frac{1}{\sqrt{2}} \rangle \rangle | 1 , 0 \rangle \rangle,
\end{align}
where $ + $ is obtained for the constant cases and $ - $ for the balanced cases. A measurement of qubit 1 distinguishes the constant and balanced cases with one evaluation of the oracle.

\begin{figure}
\scalebox{0.6}{\includegraphics{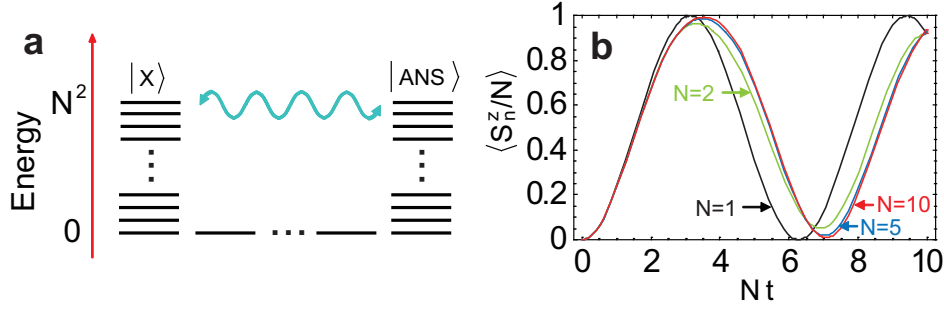}}
\caption{\label{fig3}
{\bf a} Schematic energy level structure of the Grover Hamiltonian. Rabi oscillations take place between the initial $x$-eigenstate $ |X \rangle $ and the solution state $ |ANS \rangle $.  {\bf b} Rabi oscillations executed by the Grover Hamiltonian for $ M =2 $ for various boson numbers as shown. }
\end{figure}

{\it Grover's algorithm}.  We use the continuous time formulation of the Grover search algorithm (see sec. 6.2 of Ref. \cite{nielsen00}). For the standard qubit case ($N=1$), a Hamiltonian $ H_G = |X \rangle \langle X | + | ANS \rangle \langle ANS | $ is applied to an initial state $ | X \rangle $. Here, $ | X \rangle $ is the $ \sigma^x_n = 1 $ eigenstate of all the qubits and $ |ANS \rangle $ is the solution state. Under this Hamiltonian evolution, the system executes Rabi oscillations between $ | X \rangle $ and $ |ANS \rangle $ with a period of $ t = \pi \sqrt{2^M} $ where $ M $ is the number of qubits.  

The bosonic version of the algorithm can be constructed by first writing the projection operators in $ H $ as spin operators and following the prescription as described above:
\begin{align}
H_G = N^2 \prod_{n=1}^M \frac{1}{2} \left[ 1+ \frac{S^x_n}{N} \right] + N^2 \prod_{n=1}^M \frac{1}{2} \left[ 1+ \frac{S^z_n}{N} \right], 
\end{align}
where we assumed that the solution state is $ \{ \sigma^z_n= 1 \} $ with no loss of generality.  The factor of $ N^2 $ ensures that the energy scale of all terms in the Hamiltonian are of order $ \sim N^2 $, which can be executed using the quantum bus methods as described above.  The bosonic qubits are prepared in the state 
$|X \rangle = \prod_{n=1}^M | \frac{1}{\sqrt{2}}, \frac{1}{\sqrt{2}} \rangle \rangle_n $ and evolved in time by applying $ H $. The system then executes Rabi oscillations between the initial state $ | X \rangle$ and the solution state $ | ANS \rangle $.  
The time required for reaching the solution state can be estimated from the the period of the Rabi oscillations.  For a system undergoing Rabi oscillations of a form $ \langle S^z \rangle/N = \sin^2 \omega t $, the frequency can be estimated by $ \omega = \sqrt{\frac{d^2 \langle S^z(t=0) \rangle}{dt^2}/2N} $. Evaluating the second derivative in the Heisenberg picture gives $ \frac{d^2 \langle S^z_n/N \rangle}{dt^2} = - \langle [H_G,[H_G,S^z_n/N]] \rangle = 2N^2/2^M $, corresponding to a evolution time of $ t \sim  \sqrt{2^M}/N $. The bosonic version of the algorithm has the same square root scaling with the number of sites, but with a further speedup of $ N $, resulting from the ultrafast gates made possible by the use of bosonic qubits. A numerical calculation for a simple two site case is shown in Figure \ref{fig3}b, which clearly shows the factor of $ N $ improvement in speed of the Grover algorithm. 

In conclusion, we have found that using qubits made of bosons occupying two level systems connected by a quantum bus can realize two qubit gates at a boosted energy scale of factor $ N^2 $ with no penalty on decoherence times. This allows for gate times to be reduced by a factor $ N $, resulting in many more quantum operations in a given decoherence time.  The most likely realization of such two level bosonic systems are in BECs which can be manipulated using optical transitions. Our scheme differs from previous proposals of using BECs as qubits 
in the way the states are encoded in the bosons.  Unlike previous proposals \cite{hecht04} where fragile Schrodinger cat-like states are used, the encoding presented here is robust against decoherence. For unitary rotations, there is a simple mapping connecting standard qubits to the bosonic qubit states, giving 
an equivalent operation. We leave applications including non-unitary operations such as quantum teleportation as future work.





This work is supported by the Special Coordination Funds for Promoting Science and Technology, Navy/SPAWAR Grant N66001-09-1-2024, MEXT, and the JSPS through its FIRST program. T.B. thanks R. Schmied, P. van Loock, and P. Turner for discussions.



\end{document}